\documentclass[journal]{IEEEtran}

\usepackage{cite}
\usepackage{url}
\usepackage{tabularx}
\usepackage{algorithmic,algorithm}
\usepackage{graphicx}
\usepackage{amsmath,bm}
\usepackage{amsfonts}
\usepackage{multirow}
\usepackage{booktabs}
\usepackage{hyperref}
\usepackage{xcolor}
\newcommand{\rev}[1]{#1}
\begin{document}

\title{AKDQ: Adaptive Key-Delta Quantization for Low-Bitrate RVQ-based Neural Speech Codecs}

\author{Xiao-Hang~Jiang,~Yang~Ai,~\IEEEmembership{Member,~IEEE},~Zhen-Hua~Ling,~\IEEEmembership{Senior Member,~IEEE},~Ji~Wu,~\IEEEmembership{Senior Member,~IEEE}

\thanks{This work was funded by the National Nature Science Foundation of China under Grant 62301521. (Corresponding author: Yang Ai)}
\thanks{Xiao-Hang Jiang, Yang Ai and Zhen-Hua Ling are with the National Engineering Research Center of Speech and Language Information Processing, University of Science and Technology of China, Hefei, 230027, China (e-mail: jiang\_xiaohang@mail.ustc.edu.cn, yangai@ustc.edu.cn, zhling@ustc.edu.cn).}
\thanks{Ji Wu is with the Department of Electronic Engineering, Tsinghua University, Beijing, 100084, China (e-mail: wuji\_ee@tsinghua.edu.cn).}}

\maketitle
\begin{abstract}


Low-bitrate speech coding is essential for bandwidth-constrained channels and on-device applications, but conventional residual vector quantization (RVQ)-based codecs employ a fixed number of codebooks per frame regardless of content. Adjacent speech frames often exhibit strong similarity, and quantizing them independently with the same scheme results in redundant encoding, thereby wasting bitrate.
Therefore, we propose Adaptive Key-Delta Quantization (AKDQ), which replaces the fixed per-frame VQ allocation of RVQ with an adaptive frame-wise key-or-delta policy.
AKDQ is a plug-in strategy that requires no modification to the codec's encoder--decoder backbone.
Specifically, in AKDQ, each frame is classified as either a key-frame (K-frame) quantized with all VQs, or a delta-frame (D-frame) quantized with only a few VQs on a delta correction from the previous reconstruction, thereby reducing bitrate.
The K/D decision is made by a budget-aware causal scheduler through a bit-bank mechanism, keeping the system streamable with no additional lookahead.
Experiments on both causal and non-causal codec families demonstrate that AKDQ effectively reduces bitrate redundancy.
For instance, on DAC, using AKDQ at 1.5 kbps achieves reconstruction quality comparable to using the baseline RVQ at 2.0 kbps, saving 25\% of the bitrate.

\end{abstract}


\IEEEpeerreviewmaketitle

\begin{figure*}
    \centering
    \includegraphics[width=\textwidth]{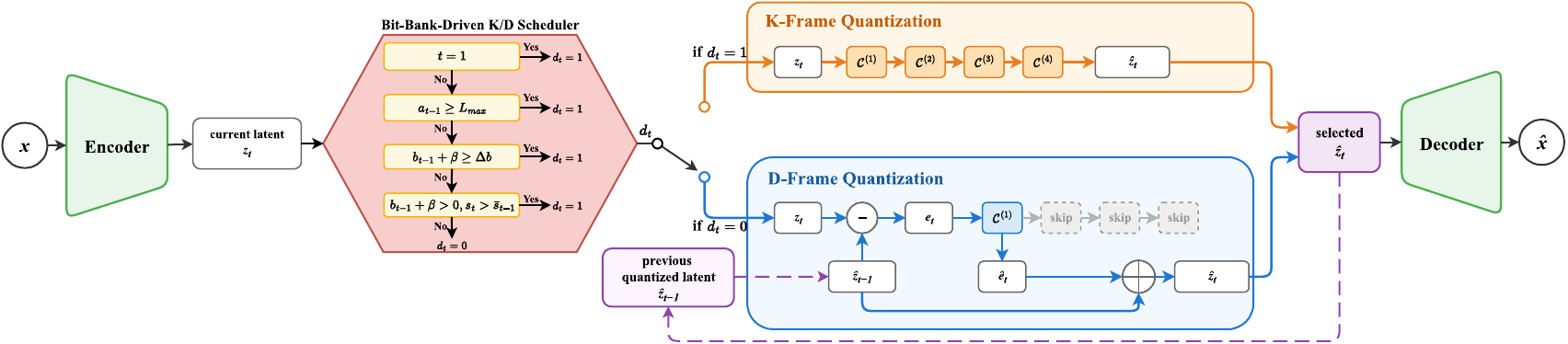}
    \caption{Overview of the proposed AKDQ with $c_K=4$ and $c_D=1$ as an example. Here, $\bm{x}$ and $\hat{\bm{x}}$ denote the speech to be encoded and the decoded speech, respectively.  $\bm{e}_t$ and $\hat{\bm{e}}_t$ denote the inter-frame residual and its quantized result, respectively. }
    \label{overview}
\end{figure*}


\section{Introduction}

\IEEEPARstart{S}{peech} codecs compress waveforms into compact bitstreams while balancing bitrate, quality, latency, and complexity~\cite{salami1994toll,brandenburg1994iso,salami1997description,keromytis2011comprehensive}.
Low-bitrate coding is especially important for bandwidth-constrained channels, on-device storage, and voice interfaces, where intelligibility and naturalness must be preserved under a tight bit budget.
Neural speech codecs have advanced the rate--distortion frontier, and many representative systems~\rev{\cite{zeghidour2021soundstream, defossez2023high, wu2023audiodec, yang2023hifi, ai2024apcodec, kumar2024high, jiang2024mdctcodec, jenrungrot2023lmcodec, zheng2025ervq, lee2026r3vq}} share an encoder--quantizer--decoder pipeline based on residual vector quantization (RVQ)~\cite{vasuki2006review}, a discrete bottleneck where later quantizers encode residuals left by earlier ones.

At low bitrates, a key limitation is that RVQ allocates the same number of codebooks to every frame regardless of content.
Consecutive speech latents are often strongly correlated~\cite{jiang2023latent}; therefore, independently quantizing every frame can redundantly transmit information already conveyed by the previous reconstruction~\cite{wang2026codecslime}.
This redundancy is particularly costly when only a few VQs are available.
A more efficient strategy would be to identify frames whose content is largely predictable from the previous quantized result and encode only the novel information, thereby avoiding the repeated transmission of similar content.

Lower bitrates can be achieved by reducing the number of active VQs, or more aggressively by adopting single-codebook or other compact quantization designs such as finite scalar quantization~\rev{\cite{mentzer2024finite,zhai2025l3ac}}.
However, preserving reconstruction quality under such reduced quantization capacity often requires modifying the encoder--decoder architecture or increasing model capacity~\cite{xin2024bigcodec,jiwavtokenizer}.
\rev{SoundStream~\cite{zeghidour2021soundstream} and EnCodec~\cite{defossez2023high} support multiple bitrates by randomly varying the active RVQ depth during training. At each operating rate, however, their codebook allocation remains uniform across frames.}
Recent variable-rate RVQ (VRVQ)~\cite{chae2025variable} adaptively allocates codebooks across frames, but it introduces an additional learnable importance subnet.
This raises a natural question: can temporal redundancy be exploited within the original RVQ bottleneck while preserving the pretrained encoder--decoder backbone and avoiding extra trainable rate-control modules?

To answer this question, this paper proposes Adaptive Key-Delta Quantization (AKDQ), which inserts a frame-wise key-or-delta transmission policy between the encoder and decoder at the RVQ bottleneck.
A \emph{key-frame} (K-frame) is quantized with all VQs and serves as a reference anchor, whereas a \emph{delta-frame} (D-frame) reuses the previous reconstruction and quantizes only a delta correction with fewer VQs.
A causal bit-bank scheduler controls the K/D decisions by tracking the surplus between the target budget and the actually consumed bits, driving the realized bitrate toward the target operating point.
Experiments on LibriTTS corpus~\cite{zen2019libritts} show that AKDQ improves the bitrate--quality trade-off over directly trained low-bitrate RVQ baselines on both causal and non-causal codec families.

\section{Proposed Method}
\label{sec:method}
\subsection{Overview}

Fig.~\ref{overview} illustrates the overview of AKDQ.
AKDQ transforms the encoder latent sequence $\bm{Z}=[\bm{z}_1,\dots,\bm{z}_T]^\top$ into reconstructed latents $\hat{\bm{Z}}=[\hat{\bm{z}}_1,\dots,\hat{\bm{z}}_T]^\top$ with a length of $T$ and produces discrete tokens for transmission.
It consists of two components: a \emph{bit-bank-driven K/D scheduler} that determines the frame type (K-frame or D-frame) for each time step, and a \emph{K/D frame quantizer} that applies the corresponding quantization strategy based on the frame type.
The transmitted discrete tokens comprise the quantized codebook indices and a frame-mode flag per frame.
During training, AKDQ is introduced by finetuning a pretrained RVQ-based codec, leaving the encoder--decoder structure unchanged and modifying only the quantization rule at the RVQ bottleneck. 
Because the K/D decision is designed to depend only on causal information, the entire pipeline of AKDQ remains streamable with no additional lookahead. 

\subsection{Bit-Bank-Driven K/D Scheduler}
\label{ssec:scheduler}

The bit-bank-driven K/D scheduler determines the frame type $d_t \in \{0, 1\}$ (K-frame for $d_t=1$, D-frame for $d_t=0$) for each time step $t=1,\dots,T$.
It must balance two objectives: (i)~insert K-frames often enough to prevent reconstruction drift caused by accumulated delta errors, and (ii)~use K-frames sparingly enough to meet the target bitrate.
We accomplish this through a bit-bank mechanism and a difficulty score.

\paragraph{Bit Bank}
We model the bitrate budget as a bank account whose accumulated deposit $b_t$ bits at time step $t$ tracks how much budget has been saved for future K-frames under the target bitrate $B_{\mathrm{target}}$.
The withdrawal bits needed to upgrade a D-frame to a K-frame are
\begin{equation}
\Delta b = (c_K - c_D)\log_2 M,
\label{eq:delta_b}
\end{equation}
where $c_K$ and $c_D$ denote the numbers of VQs used to quantize K-frames and D-frames, respectively, and $M$ denotes the codebook size. 
If every frame were a D-frame, each frame would deposit
\begin{equation}
\beta = B_{\mathrm{target}}/f_r - b_{\mathrm{flag}} - c_D\log_2 M
\label{eq:beta}
\end{equation}
bits into the bank, where $f_r$ is the frame rate, and $b_{\mathrm{flag}}$ denotes the per-frame flag bit overhead used to indicate $d_t$. 
Selecting a K-frame additionally withdraws $\Delta b$ from the bank, so the deposit evolves as
\begin{equation}
b_t =
\begin{cases}
b_{t-1} + \beta - \Delta b, & d_t=1,\\
b_{t-1} + \beta, & d_t=0,
\end{cases}
\label{eq:bit_bank_update}
\end{equation}
with $b_0 = 0$.
Intuitively, every frame deposits $\beta$ bits; whenever a K-frame is selected, $\Delta b$ is withdrawn from the bank.

\paragraph{Difficulty Score $s_t$}
Since the first frame is always assigned to a K-frame (see the initialization rule below), we set $s_1=0$ by convention.
For $t \ge 2$, we measure how well the previous reconstruction approximates the current latent:
\begin{equation}
s_t = \frac{\|\bm{z}_t - \hat{\bm{z}}_{t-1}\|_2^2}{\|\bm{z}_t\|_2^2 + \epsilon}\,,
\label{eq:frame_score}
\end{equation}
where $\epsilon$ is a small constant for numerical stability.
A large~$s_t$ indicates that the current frame deviates significantly from its reference and is therefore a strong candidate for a K-frame.
An exponential moving average $\bar{s}_t = \alpha\,s_t + (1-\alpha)\,\bar{s}_{t-1}$, initialized with $\bar{s}_2 = s_2$, tracks the running difficulty level and serves as an adaptive threshold: the scheduler compares $s_t$ against $\bar{s}_{t-1}$ so that the refresh decision is relative to the local difficulty rather than tied to a fixed threshold.

\paragraph{K/D Decision}
Let $a_{t-1}$ denote the number of consecutive D-frames since the last K-frame, counted up to and including time step $t-1$, and $L_{\max}$ cap this run length.
The scheduler applies the following priority cascade, where earlier rules override later ones.

\noindent\textbf{R1 (Initialization).} If $t=1$, the frame at time step $t$ is a K-frame to provide the first reference latent.

\noindent\textbf{R2 (Anti-drift).} If $a_{t-1} \ge L_{\max}$, the frame at time step $t$ is a K-frame to limit error accumulation in long D-frame runs.

\noindent\textbf{R3 (Sufficient deposit).} If $b_{t-1} + \beta \ge \Delta b$, the frame at time step $t$ is a K-frame because the bank can cover a K-frame withdrawal.

\noindent\textbf{R4 (Difficulty-driven).} If $b_{t-1} + \beta > 0$ and $s_t > \bar{s}_{t-1}$, the frame at time step $t$ is a K-frame to refresh a difficult frame when budget permits.

\noindent Otherwise, the frame defaults to a D-frame.
After the decision, the counter resets on a K-frame ($a_t = 0$) and increments on a D-frame ($a_t = a_{t-1} + 1$).

Over many frames, this priority cascade forms a closed-loop rate controller: deposits accumulate when D-frames dominate and are withdrawn when K-frames are selected, driving the K-frame ratio, and thus the realized bitrate, toward the value implied by $B_{\mathrm{target}}$.
Since every variable used by the scheduler is computable from the current encoder output and past state, the scheduler is fully causal and compatible with streaming inference.

\subsection{K/D Frame Quantization}
\label{ssec:residual_update}

Given the frame type $d_t$ assigned by the scheduler, the quantizer quantizes each frame along one of two paths.
A K-frame carries the full latent content without any prior reference, and is therefore quantized with all $c_K$ VQs to serve as a reference anchor.
A D-frame, in contrast, can rely on the previous reconstructed latent $\hat{\bm{z}}_{t-1}$ as a reference and only needs to quantize the inter-frame residual $\bm{z}_t - \hat{\bm{z}}_{t-1}$.
Because the residual contains less information than the full latent, it can be represented with fewer VQs ($c_D < c_K$), which saves bitrate.

Let $\mathcal{Q}_k(\cdot)$ denote quantized result obtained using the first $k$ VQs.
The two paths reuse the same RVQ codebooks $\{\mathcal{C}^{(i)}\}_{i=1}^{c_K}$, so no extra quantizer parameters are introduced.
The quantized latent is
\begin{equation}
\hat{\bm{z}}_t =
\begin{cases}
\mathcal{Q}_{c_K}(\bm{z}_t), & d_t=1 \;(\text{K-frame}),\\[4pt]
\hat{\bm{z}}_{t-1} + \mathcal{Q}_{c_D}\!\bigl(\bm{z}_t - \hat{\bm{z}}_{t-1}\bigr), & d_t=0 \;(\text{D-frame}).
\end{cases}
\label{eq:akdq_core}
\end{equation}
A K-frame costs $c_K\log_2 M$~bits while a D-frame costs only $c_D\log_2 M$~bits.
Because the scheduler routes K-frames to difficult frames and D-frames to temporally redundant ones, this is a content-adaptive allocation rather than a uniform reduction of VQs.
The reference $\hat{\bm{z}}_{t-1}$ is always the previous quantized latent regardless of its type, so it is available to both encoder and decoder without additional side information.

\subsection{Bitrate Accounting}
\label{ssec:bitrate}

Because the adaptive K/D scheduling assigns different numbers of VQs to different frames, the actual bitrate varies with the content-dependent key-frame ratio rather than being fixed by design.
Using the frame indicator $d_t$ defined in Section~\ref{ssec:scheduler}, the number of K-frames in an utterance of $T$ latent frames is $N_K = \sum_{t=1}^{T} d_t$, and the number of D-frames is $N_D = T - N_K$.
Including the per-frame flag overhead $b_{\mathrm{flag}}$, the actual bitrate is
\begin{equation}
B_{\mathrm{actual}}
= f_r\cdot\left[ \frac{(N_K\,c_K + N_D\,c_D)\log_2 M }{T}+b_{\mathrm{flag}}\,\right].
\label{eq:actual_bitrate}
\end{equation}

The target bitrate $B_{\mathrm{target}}$ should lie within a feasible range determined by the scheduler.
Let $\rho = N_K / T$ denote the key-frame ratio.
Eq.~(\ref{eq:actual_bitrate}) can be rewritten as
\begin{equation}
B(\rho) = f_r\cdot \left[\bigl(\rho\, c_K + (1-\rho)\,c_D\bigr)\log_2 M + b_{\mathrm{flag}}\right].
\end{equation}
Since the scheduler enforces at most $L_{\max}$ consecutive D-frames, $\rho$ is bounded by $1/(L_{\max}+1) \le \rho \le 1$, which yields
\begin{equation}
B\!\left(\frac{1}{L_{\max}+1}\right) \le B_{\mathrm{target}} \le B(1).
\label{eq:bitrate_range}
\end{equation}
The lower bound corresponds to the lowest K-frame density allowed by the drift constraint, while the upper bound reduces to the conventional RVQ bitrate when every frame is a K-frame.

\begin{table*}[t]
\centering
\caption{Objective \rev{and subjective} results on the 16-kHz LibriTTS test set. \rev{MOS is reported as mean $\pm$ 95\% CI half-width.} Bold and underlined results indicate the best and second-best results, respectively.}
\resizebox{0.85\textwidth}{!}{
\begin{tabular}{c  c c c c c c c c c}
\toprule
Codec & Quantizer & Bitrate (bps) & STOI $\uparrow$ & LSD $\downarrow$ & SIM $\uparrow$ & SI-SDR $\uparrow$ & ViSQOL $\uparrow$ & UTMOS $\uparrow$ & \rev{MOS $\uparrow$} \\
\midrule
\multirow{7}{*}{DAC}
& RVQ  & 1000.0 & \underline{0.893} & \underline{0.916} & 0.950 & -0.163 & 3.848 & \underline{3.395} & \rev{\underline{$3.210\pm0.555$}} \\
& VRVQ & 986.5 & 0.890 & 0.924 & \underline{0.958} & \underline{0.909} & \underline{3.895} & 3.259 & \rev{$3.140\pm0.415$} \\
& AKDQ & 1001.4 & \textbf{0.905} & \textbf{0.894} & \textbf{0.961} & \textbf{1.154} & \textbf{3.978} & \textbf{3.674} & \rev{$\mathbf{3.830\pm0.370}$} \\
\cmidrule(lr){2-10}
& RVQ  & 1500.0 & 0.913 & 0.883 & 0.967 & 2.351 & 4.075 & \underline{3.773} & \rev{\underline{$3.990\pm0.315$}} \\
& VRVQ & 1487.1 & \textbf{0.927} & \underline{0.882} & \textbf{0.975} & \textbf{3.473} & \textbf{4.182} & 3.740 & \rev{$3.900\pm0.200$} \\
& AKDQ & 1498.7 & \underline{0.926} & \textbf{0.867} & \underline{0.972} & \underline{2.571} & \underline{4.147} & \textbf{3.813} & \rev{$\mathbf{4.220\pm0.210}$} \\
\cmidrule(lr){2-10}
& RVQ  & 2000.0 & 0.935 & 0.868 & 0.975 & 3.246 & 4.183 & 3.833 & \rev{$4.330\pm0.285$} \\
\midrule
\multirow{7}{*}{MDCTCodec-S}
& RVQ  & 1000.0 & 0.870 & 0.938 & \underline{0.951} & -29.730 & 3.906 & \underline{2.549} & \rev{\underline{$2.440\pm0.260$}} \\
& VRVQ & 998.5 & \textbf{0.881} & \underline{0.933} & 0.945 & \textbf{-10.372} & \underline{3.933} & 2.448 & \rev{$2.290\pm0.275$} \\
& AKDQ & 1001.8 & \underline{0.877} & \textbf{0.927} & \textbf{0.963} & \underline{-28.497} & \textbf{4.001} & \textbf{2.581} & \rev{$\mathbf{3.060\pm0.325}$} \\
\cmidrule(lr){2-10}
& RVQ  & 1500.0 & 0.896 & 0.926 & 0.958 & -26.344 & 4.024 & 2.674 & \rev{\underline{$3.210\pm0.400$}} \\
& VRVQ & 1501.9 & \textbf{0.916} & \textbf{0.897} & \textbf{0.968} & \textbf{-6.321} & \textbf{4.238} & \textbf{3.000} & \rev{$3.020\pm0.330$} \\
& AKDQ & 1497.8 & \underline{0.904} & \underline{0.907} & \underline{0.966} & \underline{-17.514} & \underline{4.182} & \underline{2.766} & \rev{$\mathbf{3.220\pm0.225}$} \\
\cmidrule(lr){2-10}
& RVQ  & 2000.0 & 0.915 & 0.901 & 0.968 & -11.889 & 4.221 & 2.923 & \rev{$3.840\pm0.215$} \\
\bottomrule
\end{tabular}}
\label{tab:main_results}
\end{table*}

\section{Experiments and Results}

\subsection{Experimental Setup}
\label{ssec:setup}

All experiments were conducted on the 16-kHz LibriTTS corpus~\cite{zen2019libritts}, using \texttt{train-clean-100} and \texttt{train-clean-360} for training, \texttt{dev-clean} for validation, and \texttt{test-clean} for evaluation.

We evaluated AKDQ on two RVQ-based codec families, i.e., DAC~\cite{kumar2024high}, a waveform-domain codec with a non-causal convolutional backbone, and MDCTCodec-S, a causal spectral-domain codec adapted from~\cite{jiang2024mdctcodec}.
\rev{Both used improved RVQ (iRVQ) with low-dimensional, $\ell_2$-normalized lookup.}

For each codec family, AKDQ used four VQs for K-frame quantization (i.e., $c_K=4$), with a codebook size of $M=1024$ and a frame rate of $f_r=50$ Hz, while only the first VQ was used to quantize D-frames (i.e., $c_D=1$).
In the bit-bank-driven K/D scheduler, we set $L_{\max}{=}3$, $\alpha{=}0.1$, and $b_{\mathrm{flag}}{=}1$. 
With these settings, the feasible target bitrate range from Eq.~(\ref{eq:bitrate_range}) is approximately 0.925$\sim$2.05\,kbps. 
We respectively selected $B_{\mathrm{target}}=1.0$ kbps and $B_{\mathrm{target}}=1.5$ kbps for our experiments.


As baselines, we trained fixed-VQ RVQ-based codecs at 1.0, 1.5, and 2.0\,kbps for each codec family.
The 2.0\,kbps RVQ-based codecs served as upper-bound references, since AKDQ was initialized from them.
In addition, we included VRVQ~\cite{chae2025variable}, a recent variable-rate RVQ method that uses a learnable importance subnet for per-frame VQ allocation, as a comparison on both codec families.

All models were trained with AdamW \cite{loshchilov2018decoupled} on a single NVIDIA RTX 3090 GPU.
Each RVQ-based baseline was trained for 400k steps; AKDQ-based codecs branched from the corresponding 4-VQ checkpoint at 200k steps and were finetuned for another 200k steps, giving every model the same total training budget.

\subsection{Evaluation Metrics}
\label{ssec:metrics}

We evaluated reconstruction quality using six complementary objective metrics spanning different perceptual dimensions:
short-time objective intelligibility (STOI)~\cite{taal2010short}, which measures speech intelligibility preservation;
log-spectral distance (LSD)~\cite{gray2003distance}, which quantifies spectral distortion in the frequency domain;
speaker similarity (SIM)~\cite{chen2022wavlm}, which measures speaker identity preservation via cosine similarity between speaker embeddings of the original and reconstructed speech;
scale-invariant signal-to-distortion ratio (SI-SDR)~\cite{le2019sdr}, which captures time-domain waveform fidelity;
and virtual speech quality objective listener (ViSQOL)~\cite{chinen2020visqol}, which provides a reference-based perceptual quality score aligned with subjective listening.
In addition to these intrusive metrics, we also adopted UTMOS~\cite{saeki2022utmos}, a non-intrusive metric, to evaluate overall speech quality.
\rev{We also measured mean opinion scores (MOS) using a custom browser-based interface. Ten listeners rated ten utterances per condition on a five-point scale; 95\% confidence intervals (CIs) were also reported.}


\subsection{Quantization Strategy Comparison Results}
\label{ssec:baseline_results}

Table~\ref{tab:main_results} summarizes the objective \rev{and subjective} comparisons on LibriTTS.
The reported AKDQ's bitrates remained close to the intended targets on both codec families (e.g., 1001.4 and 1498.7\,bps on DAC), confirming that the bit-bank mechanism provides effective rate control without manual tuning. 
At matched bitrates, AKDQ consistently outperformed the fixed-VQ RVQ baseline across all six metrics on both codec families. \rev{AKDQ improved subjective MOS score over RVQ on both codecs at 1.0\,kbps and on DAC at 1.5\,kbps.}
The improvement is most pronounced at 1.0\,kbps, where AKDQ provided a clear perceptual-quality gain compared with RVQ (e.g., UTMOS improved by 0.28 on DAC) while also improving speaker fidelity and spectral accuracy.
This supports our motivation that, under a very small VQ budget, fixed-depth RVQ wasted bits on inter-frame redundancy, reducing effective bits and degrading speech reconstruction. 
AKDQ alleviated this through the cheaper D-frame path for temporally predictable frames. 
Compared with VRVQ at the same bitrate, AKDQ was especially strong at 1.0\,kbps, leading on all six metrics on DAC and four of six on MDCTCodec-S.
At 1.5\,kbps, the advantage over RVQ remained but became narrower, and VRVQ became more competitive, particularly on MDCTCodec-S where its learned importance subnet can exploit the larger VQ budget.

From a cross-bitrate perspective, AKDQ at 1.5 kbps approached the reconstruction quality of RVQ baseline at 2.0 kbps, i.e., saving 25\% of the bitrate.
On DAC, this is reflected in comparable perceptual quality and speaker similarity, along with a slight STOI improvement.
On MDCTCodec-S, AKDQ at 1.5 kbps still showed a certain gap in waveform fidelity, but substantially closed the perceptual-metric gap to the RVQ baseline at 2.0 kbps.
A similar pattern appears at 1.0\,kbps, where AKDQ narrowed the quality gap toward the 1.5\,kbps RVQ baseline despite using a lower bitrate.
VRVQ also benefited from adaptive allocation in this cross-bitrate comparison, especially at 1.5\,kbps, but it required an additional learnable importance module.
AKDQ instead obtained much of the bitrate-saving effect through residual delta coding and a rule-based scheduler, without modifying the encoder--decoder backbone or adding trainable components.

\begin{table}[t]
\centering
\caption{DAC ablation at 1.0\,kbps comparing AKDQ$^\ast$, AKDQ$^\dagger$, and AKDQ.}
\resizebox{\linewidth}{!}{
\begin{tabular}{l c c c c c c}
\toprule
Quantizer & STOI $\uparrow$ & LSD $\downarrow$ & SIM $\uparrow$ & SI-SDR $\uparrow$ & ViSQOL $\uparrow$ & UTMOS $\uparrow$ \\
\midrule
AKDQ               & 0.905 & 0.894 & 0.961 & 1.154  & 3.978 & 3.674 \\
AKDQ$^\ast$        & 0.723 & 1.094 & 0.898 & -8.755 & 3.131 & 1.557 \\
AKDQ$^\dagger$     & 0.899 & 0.896 & 0.955 & 0.610  & 3.898 & 3.579 \\
\bottomrule
\end{tabular}}
\label{tab:finetuning_effect}
\end{table}

\vspace{-0.5mm}
\subsection{Ablation Studies}
\label{ssec:analysis}

To validate the two core design choices of AKDQ, i.e., finetuning from a pretrained RVQ under the AKDQ forward path and using delta coding for D-frames, we conducted ablation experiments on the DAC at 1.0\,kbps, the most challenging operating point where adaptive scheduling is expected to provide the largest gains.
Table~\ref{tab:finetuning_effect} reports the results.

AKDQ$^\ast$, which directly replaced RVQ with AKDQ but skipped finetuning, suffered a substantial drop on every metric. 
For example, UTMOS fell by more than 1.8 points relative to the RVQ baseline. 
We attribute this collapse to recursive error accumulation along the D-frame chain: each D-frame conditions its quantization on $\hat{\bm{z}}_{t-1}$, which itself carries the quantization error of previous steps; without finetuning, the original RVQ codebooks have never been exposed to this error-distorted reference distribution, so reconstruction quality degrades rapidly along consecutive D-frames.
Finetuning under the AKDQ forward path realigns the codebooks with the recursive D-frame inputs and restores quality.

AKDQ$^\dagger$, which kept the scheduler but replaced the delta encoding $\hat{\bm{z}}_{t-1}+\mathcal{Q}_{c_D}(\bm{z}_t-\hat{\bm{z}}_{t-1})$ with shallow direct quantization $\mathcal{Q}_{c_D}(\bm{z}_t)$ on D-frames, recovered most of the gap but still trailed AKDQ on every metric, e.g., with noticeably lower ViSQOL and UTMOS. 
This indicates that reusing the previous quantized result as a temporal reference is essential to AKDQ's quality gains. 
It also reaffirms our core motivation that the inter-frame residual carries less information than the full latent and can therefore be faithfully quantized by a shallow quantizer, whereas applying the same shallow quantizer to the full latent (AKDQ$^\dagger$) loses too much detail.

\section{Conclusion}

In this paper, we introduced AKDQ, a plug-in quantization strategy that replaces the fixed per-frame VQ allocation in RVQ-based neural speech codecs with an adaptive key-or-delta policy, without modifying the encoder--decoder backbone. 
By leveraging the strong similarity between adjacent speech frames, AKDQ quantizes only the inter-frame residual on D-frames with fewer VQs, while reserving full-VQ quantization for K-frames that serve as reference anchors.
A bit-bank-driven causal scheduler controls the K/D assignment under a target bitrate, improving the bitrate--quality trade-off at low bitrates for both non-streaming and streaming RVQ-based codec families.
Future work will further explore broader applications of AKDQ to diverse neural speech codec architectures and more challenging ultra-low-bitrate scenarios.

\clearpage
\bibliographystyle{IEEEbib}
\bibliography{refs}

\end{document}